\documentclass[12pt]{article}
\usepackage{graphicx,wrapfig,color}
\usepackage{epsfig}
\usepackage{graphicx}
\usepackage{multicol}
\usepackage{amsfonts}
\usepackage{latexsym}
\usepackage{amsmath,amssymb}
\usepackage{enumerate}
\usepackage{enumitem}
\numberwithin{equation}{section}
\usepackage{cite}

\newcommand{\ben}{\begin{enumerate}}
\newcommand{\een}{\end{enumerate}}
\newcommand{\spa}{\phantom{s}}
\newcommand{\bea}{\begin{eqnarray}}
\newcommand{\eea}{\end{eqnarray}}
\newcommand{\be}{\begin{equation}}
\def\bel#1{\begin{equation} \label{#1}}
\newcommand{\ee}{\end{equation}}
\newcommand{\bi}{\begin{itemize}}
\newcommand{\ei}{\end{itemize}}
\newcommand{\ba}{\begin{align}}
\newcommand{\ea}{\end{align}}
\newcommand{\comments}[1]{}

\def\pref#1{(\ref{#1})}
\def\hf{\frac12}
\def\del{\partial}

\def\vp{\varphi}
\begin{document}

\begin{titlepage}

\hfill{HRI/ST 1412}

\vskip 1.0 cm
\begin{center}
{\Large \bf  
Inflationary Constraints on\\ Late Time Modulus Dominated Cosmology
}
\vskip 1.0 cm  
{ 
{\sc{Koushik Dutta}}$^{*}$ and {\sc{Anshuman Maharana}$^{\dagger}$}
\let\thefootnote\relax\footnotetext{Email: {$^{*}$$\mathtt{koushik.dutta@saha.ac.in}$,   {$^{\dagger}$$\mathtt{anshumanmaharana@hri.res.in}$ }}}
}

\vskip 0.2 cm
{\textsl{$^{*}$Theory Division \\
 Saha Institute of Nuclear Physics \\
 1/AF Salt Lake  \\
 Kolkata, 700064, India}
}
\vskip 0.2 cm
{\textsl{$^{\dagger}$Harish Chandra Research Intitute \\
Chattnag Road, Jhunsi\\
Allahabad, 211019, India\\}
}
\end{center}

\vskip 0.1cm

\begin{abstract}
\large{We consider cosmological scenarios in which density perturbations are generated
by the quantum fluctuations of the inflaton field at early times; the late time dynamics involves
a modulus which first dominates the energy density of the universe and
then decays to reheat the visible sector.  By examining the evolution of
energy density of the universe from the time of horizon  exit of a pivot mode to the present day,
and the fact that a modulus field decays via Planck
suppressed interactions, we arrive at a relation which relates the mass
of the modulus, inflationary observables/parameters and broad characteristics of the post inflationary reheating phase. When viewed together with  generic
expectations regarding
reheating  and 
the initial field displacement of the modulus after inflation, the relation gives a bound on
the minimum mass of the modulus. 
For a large class of models, the bounds obtained (for the central values of
 Planck data) can be much stronger than the  
``cosmological moduli problem" bound.}
\noindent

\end{abstract}

\vspace{2.0cm}

\end{titlepage}
\pagestyle{plain}
\setcounter{page}{1}
\newcounter{bean}
\baselineskip18pt
%
%
\section{Introduction}

The hot big bang model together with the inflationary paradigm provides a highly
attractive framework for cosmology.  Typically, it is assumed that after inflation
the visible sector  degrees of freedom reheat and  have evolved adiabatically since then. Baryon asymmetry  is tied to high scale physics via mechanisms such as leptogenesis and dark matter is a thermal relic. In spite of the impressive successes of this scenario it is important to keep in mind that the big bang model is consistent with entropy production as long as this took place before nucleosynthesis.

     In fact, the cosmology in many beyond the standard model scenarios can potentially involve late time entropy production as a result of decay of long lived light scalar fields (light moduli in the context of supergravity/string models). Light moduli are typically displaced from their minimum during inflation. At the end of inflation the universe reheats,  energy associated with the inflaton gets converted to radiation. As the universe expands the Hubble constant decreases, when it becomes of the order of the mass of a modulus the modulus begins to oscillate
about the minimum of its potential. Subsequently, the energy density associated with the field begins to redshift
like matter, at a rate significantly slower than the radiation -- the energy  associated with the modulus can quickly dominate the energy density of the universe. Eventually the modulus decays reheating the universe.

The last modulus to decay\footnote{
The non thermal matter dominated universe must end prior to  big bang nucleosynthesis. We know with great confidence that at the time of primordial nucleosynthesis the universe was radiation dominated. } essentially provides the ``initial conditions" for cosmological evolution. The reheat temperature is given by the decay width of the scalar field. The properties such as dark matter density, baryon asymmetry are determined by the  branching ratio of the various decay products. This makes the scenario highly predictive for  models where it is possible to compute the couplings of the scalar field to the standard model and other light degrees of freedom. This predictivity comes at a cost; the late time reheating typically washes out a lot of  information making the connection to  early universe physics very challenging.

  In this paper we will obtain a relationship between the mass of the modulus, inflationary observables and broad characteristics of post inflationary reheating (number of e-foldings and the effective equation state). When viewed along with the usual expectations
regarding the equation of state during reheating  and the initial field displacement of the
modulus,  this  immediately leads us to a lower bound on the mass of the modulus; cf. Eq.~\eqref{bond}. 
For a large class of inflationary potentials, the bounds can be stronger than those provided by the ``cosmological moduli problem".   The constraint depends exponentially on the number of e-foldings between horizon exit and the end of inflation, future observations  are going to be crucial in sharpening the limits.    

This paper is organised as follows. In section \ref{cosrel} we discuss some basic aspects 
of  the dynamics  of cosmologically relevant scalars. We also briefly discuss the motivations for their appearance in string/supergravity models and some related phenomenological scenarios.  In section \ref{ourana},  we present the our analysis, obtain the bound and discuss  implications. We conclude in section \ref{conclu}.  
   
\section{Cosmologically Relevant Scalars}
\label{cosrel}

As discussed in the introduction, long lived scalar fields with masses below the Hubble scale 
during inflation are expected to be cosmologically relevant. Here we give a brief review of  some
aspects that will be relevant for us and refer the reader to the seminal papers \cite{cmp,ccmp,cmmp}
for a complete discussion.

   The cosmological evolution of a scalar field is given by
\bel{scalarevol}
 \ddot{\varphi} + (3H + \Gamma_{\varphi}) \dot{\varphi} +  { \del V (\varphi) \over \del \varphi} = 0,
\ee
where $H$ is the Hubble constant and $\Gamma_{\varphi}$ the width of the scalar. For long lived scalars,
this implies  that the field is frozen at its initial displacement $\varphi_{\rm in}$ if $H > m_{\varphi}$ . The 
initial displacement can be due to thermal/quantum fluctuations of the field during inflation \cite{quan} or explicit dependence of the potential for the modulus on the inflaton vev \cite{mismatch, mismatchb, DineR, DineRT} and is expected to be of the order of $M_{\rm pl}$. After the modulus begins to oscillate about its minimum; matter radiation equality takes place at total
energy density
\bel{equalrhho}
    \rho_{\rm eq} = m^{2}_{\varphi} \vp_{\rm in}^2 \bigg( {{  \vp^2_{\rm in}} \over 6 M_{\rm pl}^2} \bigg)^3.
\ee
The modulus then dominates  the energy density of the universe. It decays at time $t = \tau_{\rm mod}$, the corresponding energy density is
\bel{decayden}
   \rho_{\rm decay} \sim M_{\rm pl}^2 \Gamma_{\varphi}^2,
\ee
where $\Gamma_{\varphi}$ is the total width of the field. Moduli fields interact only via Planck suppressed interactions, their lifetime
\bel{width}
   \tau_{\rm mod} \approx {1 \over \Gamma_{\varphi} } \approx   { 16 \pi M_{\rm pl}^2 \over  m_{\varphi}^{3} }.
\ee
Combining \pref{decayden} and \pref{width} one obtains the reheat temperature in terms of the mass 
\bel{rehit}
         T_{\rm rh2} \sim m_{\varphi}^{3/2}M_{\rm pl}^{-1/2}.
\ee
For successful nucleosynthesis one requires
the reheating temperature to be greater than a few MeV, and 
one obtains the famous cosmological moduli problem bound \cite{cmp,ccmp,cmmp}
\bel{bound}
    m_{\varphi} \gtrsim 30 \spa {\rm TeV} .
\ee
More detailed cosmological bounds on late time entropy production have been calculated in \cite{Kawasaki:1999na}.

 Next, we briefly mention the arguments
which suggest that cosmologicaly relevant moduli fields can be  expected to be present in supergravity/string constructions  with gravity mediation of supersymmetry  breaking (though our analysis in no way  commits  to these models,
the bound obtained is equally valid if all the moduli are stabilised supersymetrically). The  form of the supergravity
F-term potential
\bel{pot}
   V = e^{{\cal{K}}} \bigg( {\cal{K}}^{ij} D_i W D_j W  -  3 |W|^2 \bigg)
\ee
together with the formula for the gravitino mass
\bel{gravmass}
  m_{3/2} = e^{{\cal K}/2} |W|
\ee
suggests that once supersymmetry is broken the moduli receive a contribution to their potential which is of the
order of the gravitino mass. Furthermore, in gravity mediated  supersymmetry breaking the scale of the
soft masses is of the order of the gravitino mass. Though there can be  exceptions
to both these expectations  it is reasonable to expect moduli masses at the 
supersymmetry breaking scale or at most few orders of magnitude higher. For models of low energy supersymmetry,
 this implies moduli masses at the TeV scale or few orders of magnitude
higher in the context of gravity mediated models.  There is a tension between the bound \pref{bound} and supersymmetry as a solution to the hierarchy problem; though sequestering
can alleviate the problem.

   Recently, cosmology with a modulus/moduli decaying at late times has emerged as the ``preferred scenario"
in many string constructions. There has been substantial exploration of the associated phenomenology
in the context of M-theory compactifications \cite{mdm, mdm2}. In these models, the non-thermal dark matter density \cite{randall} is of the right order of magnitude if the moduli are at 30-100 TeV. Based on this,  the expectation that supersymmetry provides a partial resolution of the hierarchy problem\footnote{With soft scalar masses at 100 TeV, one
can attribute 14 orders of magnitude of the hierarchy problem (of the total 16) to supersymmetry.} and other phenomenological
successes \cite{piyushwa} soft masses at 30-100 TeV have been suggested to be a generic prediction of string compactifications
\cite{bobby, douglas}.

  The Large Volume Scenario \cite{lvs1, lvss1} in type II B provides an explicit realisation of moduli stabilisation. In these
models there is one modulus (corresponding to the overall volume of the compactification) which is significantly lighter than all others - it is the  last to decay. The models can exhibit sequestering \cite{tseque1, tseque2, tseque3}, there
is no theoretical tension with TeV scale supersymmetry. A generic prediction of the cosmology is  dark radiation in the form of axions \cite{darkr,darkrr}. In addition, the scenario provides non thermal generation of dark matter \cite{Allahverdi}, and its relation to the baryon energy density \cite{Allahverdi:2010rh}. For more phenomenological issues, see \cite{Fan:2014zua}.

\section{Inflationary Constraints on Light Moduli}
\label{ourana}

In this section we obtain a relationship between
the mass of the late decaying modulus field, post-inflationary reheating parameters
and inflationary observables. This will involve two steps. First, we  obtain
an expression for the number of e-foldings in which the universe is matter (modulus) dominated
which follows from the evolution of the energy density of the universe from the time of
horizon exit to the present day\footnote{This can be considered as the generalisation (for late time modulus dominated
cosmology) of
the equation which gives the total number of e-foldings  (see for e.g. \cite{planck}) in inflationary models.}. Then, we express the number of e-foldings in terms of the
modulus mass by taking its lifetime to be as given by Eq.~\pref{width}.

Our working assumption will be that the inflaton is responsible for the observed density perturbations; more specifically, none of the light moduli play the role of a curvaton. The history of the universe will be taken to be as described in the introduction and section \ref{cosrel} 
where reheating phase produced by the inflaton decay is followed by a prolonged matter dominated phase from the coherent oscillation of the modulus field. This scalar field decays before the epoch of big bang nucleosynthesis. We will be explicitly including
only one modulus  (the one to decay last) in the history. 

    We characterise the reheating phase  by two parameters - the number
of e-foldings during the era $N_{\rm re}$, and the equation of state $w_{\rm re}$. Given this
general parametrisation we hope to capture not only the transfer of energy from the inflaton
to radiation but also of the decay of other moduli (if they are significantly heavier, and
decay much earlier) by the ``reheating'' phase.

   We begin the derivation of our relation by writing the condition which determines the
exit of a mode of comoving wavenumber k from the horizon  $k = a_{\rm k} H_{\rm k}$  as
\bel{fe}
  k = {a_{\rm k} \over a_{\rm end} }.  {a_{\rm end} \over a_{\rm re} }.{a_{\rm re}  \over a_{\rm eq} }. 
  {a_{\rm eq}  \over a_{\rm decay} } .a_{\rm decay} H_k,
\ee
where the subscripts end, re, eq and decay indicate the end of inflation, end of reheating, equality
of energy density between matter (in this case modulus energy density) and radiation, and decay of the modulus. Taking the logarithm of
\pref{fe} one obtains
\bel{1kinet}
 N_{\rm matdom}  = - N_{\rm k} - N_{\rm re} - N_{\rm rad} - \ln k + \ln(a_{\rm decay}) + \ln H_k
\ee
where $N_{\rm matdom}$ is the number  of e-foldings in the matter (modulus) dominated  era, $N_{\rm k}$ the number of e-foldings between
the horizon exit of the mode with mode number k and end of inflation,  $N_{\rm re}$
the number of e-foldings during the period of reheating and $N_{\rm rad}$ the number of e-foldings in the 
radiation dominated era.


  Next, we obtain another expression for $N_{\rm matdom}$ based on the evolution of energy
density. We  begin by writing\footnote{The energy density $\rho$ with a superscript will denote the energy density in
a given form; $\rho^{\rm matter}_{\rm decay}$ is the energy density in the form of matter at the time
of modulus decay.}
\bel{nmatt}
    - N_{\rm matdom} =  {1 \over 3 } \ln \big( { \rho^{\rm matter}_{\rm decay}/ \rho^{\rm matter}_{\rm eq} } \big).
\ee
The energy density at the time of decay can be expressed in terms of the reheat temperature $T_{\rm rh2}$
and the effective number of light species $g_{\rm rh2}$ at the time of reheating\footnote{
We will explicitly incorporate the phase of final reheating later and find that conclusions are not altered.} 
\bel{ed}
 \rho^{\rm matter}_{\rm decay} \approx \rho_{\rm decay} = (\pi^{2}/30) g_{\rm rh2} T^4_{\rm rh2},
\ee
and the reheat temperature can be related to the CBM temperature today  (assuming dark radiation is insignificant) by
\bel{trel1}
 \spa \spa   \spa \spa T_{\rm rh2} = \left( {43/ 11 g_{\rm s, rh2} } \right)^{1/3} \left( {a_0/ a_{\rm decay} }\right) T_0
\ee
where $g_{\rm s, rh2}$ is the effective number of light species for entropy. Combining Eq.~\eqref{ed} and Eq.~\eqref{trel1}
and expressing $\ln \rho_{\rm eq}^{\rm matter}$  as
\be
\label{rhodivt}
  \ln( \rho^{\rm matter}_{\rm eq} ) = \ln \left({ \rho^{\rm radiation}_{\rm eq} / \rho_{\rm re} }\right) + \ln \left({ \rho_{\rm re}/ \rho_{\rm end} }\right) +\ln (\rho_{\rm end}), 
\ee
Eq.~\eqref{nmatt} yields (we take $\rho_{\rm re}^{\rm radiation} \simeq \rho_{\rm re}$)
\bea
\label{eq10t}
  - {3 \over 4} N_{\rm matdom} &=&{1 \over 4} \ln \left(\pi^{2}g_{\rm rh2}/30  \right) + {1 \over 3} \ln \left( 
  {43/11 g_{\rm s, rh2} } \right)  + \ln \left( {a_0 T_0/ a_{\rm decay} } \right) + N_{\rm rad} \cr 
    &\phantom{=}& -{1 \over 4}\ln ( \rho_{\rm end} ) +  {3\over 4}(1+w_{\rm re}) N_{\rm re}.
\eea
Adding  \pref{1kinet} and \pref{eq10t} the dependence on both $N_{\rm rad}$ and $a_{\rm decay}$ drops out and we 
obtain 
\bea
\label{eq11t}
{ 1 \over 4} N_{\rm matdom} &=& {1 \over 4} \ln \left(\pi^{2}g_{\rm rh2}/30\right) + {1 \over 3} \ln \left( 
  {43/ 11 g_{\rm s, rh2} } \right)  - N_k - \ln \left( {k/ a_0 T_0 } \right) + \ln H_k \cr
  &-& {1 \over 4}\ln ( \rho_{\rm end} ) - { 1 \over 4} ( 1 - 3w_{\rm re}) N_{\rm re}
\eea

Finally, we express $N_{\rm matdom}$ in terms of the modulus mass and lifetime by using the explicit
form of the scale factor as a function of time. Recall that if the equation of state is $w$ the evolution of the scale factor between
times $t_1$ and $t_2$ is given by
\bel{wevol}
    \left( {a (t_2)/  a (t_1) }\right)^{\frac{3}{2}(1+w)} =  1+ {3 \over 2} (1+w){H}(t_1)  (t_2 - t_1).
\ee
By demanding that the time elapsed between the end of inflation and the decay of the modulus is
the lifetime of the modulus (and assuming  $N_{\rm matdom} \gg 1; \spa N_{\rm re}, N_{\rm eq} > 1$), we 
obtian\footnote{We will approximate the evolution by including  (in the right
hand side Einstein equation) only 
the dominant component of the energy density in each epoch. We hope to
report the results of an exact treatment based on numerics in the future.}
\bea
\label{decaytime}
   \bigg( {a (t_{\rm decay}) \over  a ({t}_{\rm eq}) }\bigg)^{3/2} &=& { 3 \over 2} H_{\rm eq}\tau_{\rm mod}
   - {3 H_{\rm eq} \over 4 H_{\rm re}  } \bigg( {a (t_{\rm eq}) \over  a ({t}_{\rm re}) }\bigg)^{2} -  {H_{\rm eq} \over  (1+w)H_{\rm end}  } \bigg( {a (t_{\rm re}) \over  a ({t}_{\rm end}) }\bigg)^{(3/2)(1+w)}  \cr
 &=&  {3 \over 2}H_{\rm eq}\tau_{\rm mod} -{3 \over 4} - {1 \over 1 + w} e^{-2N_{\rm rad} } \approx {3 \over 2} H_{\rm eq} \tau_{\rm mod} 
\eea
Thus,
\bel{dcc}
  N_{\rm matdom} \approx {2 \over 3} \log (3/2) + {2 \over 3} \log (H_{\rm eq} \tau_{\rm mod} )
\ee
It can easily be checked that the above (approximate) expression is also correct in the
regime $N_{\rm matdom} \gg 1$ and $N_{\rm re}, N_{\rm rad} \ll 1$. Next, we substitute for
$H_{\rm eq}$ by using Eq.~\eqref{equalrhho} and parametrise the initial displacement
as $\varphi_{\rm in} = Y M_{\rm pl} $ to obtain  

\bel{nnmmat}
  N_{\rm matdom} = - {2 \over 3} \ln 3 - {5 \over 3} \ln 2 + {2 \over 3} \ln m_{\varphi} \tau
                             + {8 \over 3} \ln Y.
\ee
 Equating the  two expressions for $N_{\rm matdom}$ given by  \pref{eq11t} and \pref{nnmmat}; and making
use of the slow roll expression for the  Hubble constant, $H_k^2 = \frac{1}{2}\pi^2 M_{\rm pl}^2 r A_s$
(with $A_s$ the amplitude of scalar fluctuations and $r$ the tensor to scalar ratio) we get
\bea
\label{general}
 {1 \over 6} \ln m_{\varphi}  \tau_{\rm mod} + {1 \over 4}( 1 - 3w_{\rm re}) N_{\rm re} + {2 \over 3 } \ln Y &=& {1 \over 4} \ln \left(\pi^{2} g_{\rm rh2}/30  \right) + {1 \over 3} \ln \left( 
  {43/ 11 g_{\rm s, rh2} } \right)\cr &-& \ln \left( {k/ a_0 T_0 } \right)  + \frac{1}{12}\ln(4/3)\cr  &-& { 1 \over 4 } \ln \left({ \rho_{\rm end}}/ \rho_{\rm k} \right) + {1 \over 4} \ln { (\pi^{2} r A_s) } - N_k. 
\eea
 We would like to reiterate that so far no assumption about inflationary physics has been made except for slow-roll.  All the inflationary details are encoded in the last three terms of the right hand side of the above equation. We  use  
 {\small{PLANCK}} data \cite{planck} for quantities that have already been observed with accuracy; the primordial scalar amplitude $\ln(10^{10} A_s) = 3.089$ at  the pivot scale $k = 0.05 \spa {\rm{Mpc}^{-1}}$ and $T_{0}=2.725 \spa K$. For the number
of degrees of freedom\footnote{The dependence on the number of degrees of freedom appears as 
 $\ln \left( { g_{\rm rh2}^{1/4} \big{/} g_{\rm s,rh2}^{1/3}} \right)$, hence is quite mild.}, we use $g_{\rm rh2} \approx g_{\rm s,rh2} \approx 100$.   Moreover, we take $\tau_{\rm mod}$ for the modulus as given by \eqref{width}.

 Plugging in all this, we find
 \bea
\label{general1}
 {1 \over 6} \ln \left(\frac{16 \pi M_{Pl}^2}{m_{\varphi}^2}\right) + {1 \over 4}( 1 - 3w_{\rm re}) N_{\rm re} + {2 \over 3 } \ln Y = 55.43 +\frac{1}{4}\ln r - N_k   - { 1 \over 4 } \ln \left({ \rho_{\rm end}}/ \rho_{\rm k} \right),  
\eea
The above equation is our main result. This equation is the generalisation of the equation giving the number of $e$-foldings between horizon exit for the modes relevant for CMB observations and the end of inflation. It can be used to obtain the preferred values of $N_k$ given the lightest modulus mass.\footnote{ We note that for $m_{\varphi} \approx 100 \spa 
\rm{TeV}$ and a typical value of Y ($Y \approx 1/10$);  the central value of $N_k \approx 45$.}

 There are strong reasons
to believe (supported by both analytic and numerical work) that the equation of state during reheating satisfies $w_{\rm re} < 1/3$ (see for e.g. \cite{reheat, planck} for  discussions). From
the point of a scalar field $\chi$ oscillating about its minimum $w_{\rm re} > 1/3$ corresponds to the scalar field potential
near its minimum being dominated by higher dimensional operators (greater that $\chi^{6}$) hence can be considered
unnatural\footnote{Although $w_{\rm re} > 1/3$ cannot be excluded, \cite{except} provided a model.}. Guided by this
we take $w_{\rm re} < 1/3$  -- the second term in the left hand side is positive definite. Now, the above equation can be easily converted to an lower limit for the mass of the modulus\footnote{We note that if the reheating phase is
almost instantaneous i.e $N_{\rm matdom} \gg N_{\rm re}$ then the condition $w_{\rm re} < 1/3$ is not necessary.}
 \bea
 \label{bond}
 m_{\varphi}\gtrsim   \sqrt{16 \pi} M_{\rm pl} Y^2 ~e^{-3\left(55.43  - N_k   + { 1 \over 4 } \ln \left({ \rho_{\rm k} / \rho_{\rm end} } \right)+\frac{1}{4}\ln r\right)}.
 \eea

We note that even if  we consider highly exotic reheating $w_{\rm re} > 1/3$, \pref{general} predicts values $m_{\varphi}$ to be quite large
for $N_k \approx 50$; as long as the number of e-foldings during reheating is
not comparable to the number of e-folding of modulus domination (as is expected for a light modulus).
We will discuss this later and focus on \pref{bond} for now. Recall also that Y is the initial field displacement of the light modulus in Planck units. As discussed
in section \ref{cosrel} the generic expectation for the initial displacement  is of the order of $M_{\rm pl}$,
it cannot affect the value of the right hand side of \pref{bond} significantly. We note in passing that in deriving the CMP bound one also makes use of the fact that Y is not expected to be significantly less that one.
We emphasise that we have been  conservative in  estimating the bound; a long reheating phase would make it stronger. When the final reheating phase from the decay of the modulus field is carefully considered in this picture, an extra term similar to the 2nd term in Eq.~\eqref{general1} appears that involves $w_{\rm re2}$ and $N_{\rm re2}$ of the final reheating epoch. Following similar arguments as outlined above, the bound remains unchanged.


We would like to briefly comment
on the multiple modulus case. As mentioned earlier, given the general parametrisation of the reheating phase,
the dynamics of heavier moduli which decay very early on should be captured in the ``reheating phase". The relevant 
dynamics involves epochs of matter domination and radiation domination, should satisfy the bound $w_{\rm re} < 1/3$.
If there are $N$ moduli at the same mass scale (with a diagonal Kahler metric or if we make the assumption that
the  Kahler metric is generic as in \cite{many}) then the
 energy density at equality \pref{equalrhho} scales as $N^4$ (for fixed $\varphi_{\rm in}$); the bound becomes stronger by a factor of $N$.

Before studying the bound in the context of various inflationary scenarios, we make a few comments
\bi
\item Larger the number of e-foldings,  stronger the bound. Typically
\bel{nsnk}
    N_{\rm k}  \approx   { \beta \over  {1 - n_s}}
\ee
with $\beta$ a model dependent constant \cite{talk, Burgess:2013sla}.  The bound is highly sensitive to the spectral tilt.

\item The second parameter in the exponent, ${ 1 \over 4 } \ln \left({ \rho_{\rm k} / \rho_{\rm end} } \right)$
is positive definite. A large ratio between the energy density at the time of horizon exit and the end of inflation weakens the bound. 
\item The third parameter ${1 \over 4} \ln r$ is negative definite. By itself, this term would strengthen the bound
for low values $r$.
\ei
One should be careful while using the above points to estimate the bound.  Given an inflationary model
the three parameters in the exponent will not be independent of each other. But as we will see next, without committing much to a particular model of inflation, we can obtain a good understanding based on the class of models. 
\subsection{Small field models}

In this class of models, the field variation is sub-Planckian in a typical plateau like potential. The change in energy density between  horizon exit  and end of inflation
is not significant  for this kind of models. In this case, it is reasonable to drop the term involving the logarithm of the two energy densities in the exponent of \pref{bond}. Taking $r = 0.01$,  we get
 \bea
 \label{bond1}
 m_{\varphi} \gtrsim   \sqrt{16 \pi} M_{\rm pl}Y^2 {\huge{e}}^{ -3\big( 54.28 -    N_{\rm k} \big) }
 \eea
To get a feel for the numbers, $N_{\rm k} = 50$  and taking $Y = 1/10$ (in what follows, we will always take Y = 1/10 while quoting numbers) we get
\bel{number1}
   m_{\varphi} \gtrsim  4.5 \times 10^{8} \spa \rm{TeV}
\ee
which is well above the  bound \pref{bound} given by the cosmological moduli problem.

   Now, we comment on the case of non-generic reheating ($w_{\rm re} > 1/3$). A useful parametrisation of the length of reheating phase can  be done by defining a constant $\lambda$ by $N_{\rm re} (1 - 3w_{\rm re}) \approx - \lambda N_{\rm matdom}$. Even if we take a long exotic reheating phase with $\lambda \approx 1/3$, a direct application of
 \pref{general1} gives 
 (for $N_{k} \approx 50$) the mass to be
\bel{nega}
   m_{\varphi} \approx  10^{6} \spa \rm{TeV}.
\ee
This is again well above the cosmological moduli problem bound.

   Given an inflationary model, $N_{k}$ can be determined from precise measurements of $n_s$ via
a relation of the  form \pref{nsnk}.  A survey of the values
of $\beta$  associated with various models is given in \cite{talk, Burgess:2013sla}.  For a typical value models $\beta \approx 2$, and $n_s=0.9603$ as given by the central value of {\small{PLANCK}}; $N_k \approx 50$. The bound in this case is
$4.5 \times 10^{8} \spa \rm{TeV}$. Of course, given the form of \pref{nsnk} a small variation
in $n_s$ can lead to an large change in $N_{k}$; this can alter the bound significantly given the exponential dependence.  At $1 \sigma$,  the variation of the spectral index for {\small{PLANCK}} is $\Delta_{1 \sigma} n_s = 0.0073$. The $1 \sigma$ upper  limit of $n_s$ gives  $m_{\varphi} > M_{\rm pl}$, ruling out late time modulus cosmology (for small field models with $\beta =2$,  $r =0.01$).
 On the other hand the lower value gives  $m_{\varphi} \gtrsim 0.1 \spa {\rm TeV}$,   two to three orders of magnitude below the  
 one given by the cosmological moduli problem. Given this sensitivity\footnote{Theoretically, the exponential sensitivity  implies that ${1 /N_k}$ corrections to \pref{nsnk} can be relevant in some models. We have computed the leading corrections to \pref{tilt} for monomial potentials and found that they do not significantly alter the bound.}, future experiments\cite{euclid, prism, Ko, BBO} which will lower the uncertainties in the measurement of $n_s$ by one order of magnitude  will play an important role (the uncertainty in the mass will reduced to two orders of magnitude) in determining the existence of late time modulus decay cosmology in this context.

\subsection{Large Field Models}
                      As prototypes of the large fields models we consider models with where the 
inflation potential is a monomial $V_{\chi} = \hf m^{4 - \alpha} \chi^\alpha$ (keeping the``$m^{2} \chi^{2}$" model \cite{Piran, Belin, Linde} and axion monodromy  \cite{monodromy, west} models in mind). 
For these models
\bel{tilt}
  n_s -1  = -(2 + \alpha)/(2N_{k}), \spa \spa  r = 4 \alpha/N_{k}.
\ee
Thus measurement of  $n_s$ and $r$ determines both $N_k$ and $\alpha$.  But given the uncertainty in
the measurements of $r$, we will take observational input only from  $n_s$. We will treat $\alpha$
as a model building parameter -- $N_{\rm k}$ and $r$ will be treated as the derived quantities in \pref{tilt}.

 Let us evaluate the exponent in \pref{bond}. From \pref{tilt} we have  $N_{\rm k} = (2 + \alpha)/2(1 - n_s), r =  8 \alpha (1-n_s)/ (\alpha +2)$. To compute  $\log(\rho_{\rm k} / \rho_{\rm end})$ we need to express the energy densities at the time of horizon exit and end of inflation in terms of $n_s$ and $\alpha$. The energy density at
the time of horizon  exit is simply the value of the of the potential at the time of horizon exit, $\rho_k =
\hf m^{4 - \alpha} \chi_{\rm k}^{\alpha} $. The displacement at the time of horizon exit is given by $\chi_k^2 = 2 \alpha M^2_{\rm pl} N_k$. On the other hand,  at the end of inflation the energy density
is given by $\rho_{\rm end} = (1  + \lambda) V_{\rm end} = \hf (1 + \lambda ) m^{4 - \alpha} \chi_{\rm end}^{\alpha}$; where $\lambda = \big( 1  + { 2 /\epsilon_0} \big)^{-1}$ with $\epsilon_0$ the value
of the slow roll parameter $\epsilon$ at the end inflation $(\epsilon_0 \approx 1)$. The field displacement 
at the end of inflation is given by $\chi_{\rm end}^2 =  (\alpha^2 M_{\rm pl}^2/ 2 \epsilon_0)$. 
Combining the above, we find
\bel{chaos}
 - N_{\rm k} +  { 1 \over 4 } \ln \left({ \rho_{\rm k} / \rho_{\rm end} } \right)+ \frac{1}{4}\ln r = 
- {(2 + \alpha) \over 2(1 - n_s) }  -{1 \over 4} \ln 3 + {1 \over 8} (\alpha +8) \ln 2  + { 1 \over 8} (\alpha -2) \ln \bigg(  { { 2 + \alpha}  \over \alpha (1 - n_s)} \bigg);
\ee
the bound becomes
 \bea
 \label{bond2}
 m_{\varphi} \gtrsim   \sqrt{16 \pi} M_{\rm pl}Y^2 {\huge{e}}^{ -3 \big(  55.85 -  {(2 + \alpha) \over 2(1 - n_s) }  + {\alpha \over 8}  \ln 2  + { 1 \over 8} (\alpha -2) \ln \big(  { { 2 + \alpha}  \over \alpha (1 - n_s)} \big)   \big)}
 \eea
 The coefficients  of the logarithmic terms are such that not much error is made if one drops the
terms; we write
 \bea
 \label{bond3}
 m_{\varphi} \gtrsim   \sqrt{16 \pi} M_{\rm pl}Y^2 {\huge{e}}^{ -3\big( 55.85 -    {(2 + \alpha) \over 2(1 - n_s) } \big) }
 \eea
  For the $\hf m^{2} \chi^2$ and $n_s$ as given by the central value of {\small{PLANCK}} the bound
is $m_{\varphi} \gtrsim  10^{7} \spa \rm{TeV}$, one order of magnitude below \pref{number1}. On the other hand for $\alpha =1$ and $n_s$ at
the same value the bound becomes $m_{\varphi} > 10^{-10} \spa \rm{TeV}$, which is completely irrelevant.

To summarise, for $N_{k} \approx 50$, the bound is significantly stronger than that provided by
the cosmological moduli problem, and larger the $N_{k}$, stronger the bound is. From the point of view of microscopic
models, there is a high sensitivity on the parameters in the inflaton potential as a result of the exponential dependence on $N_{k}$ and
the fact that $N_{k}$ scales as the inverse of the spectral tilt.  The parameter $\beta$ (as defined in \pref{nsnk}) plays
a central role in determining the magnitude of the bound; $\beta \approx 2$ (which seems to be the
typical value, see  for e.g. \cite{talk}) and $n_s$ at the central value of {\small{PLANCK}} correspond
to $N_{k} \approx 50$ and hence imply a strong bound.

 For the non-generic case of exotic reheating with $w_{\rm re} > 1/3$, it is useful to parametrise the duration of
reheating by a parameter $\lambda$; $N_{\rm re} (1 - 3w_{\rm re}) =  - \lambda N_{\rm matdom}$. Even for $\lambda
= 1/3$ (which corresponds to a rather long phase of exotic reheating)    direct use of \pref{general} gives  (for $N_{k} \approx 50$) $m_{\varphi} \approx    10^{6} \spa \rm{TeV}$. Again, well above the CMP bound.

      As discussed earlier,  we have   taken moduli interactions to be
 Planck suppressed in obtaining  \pref{bond}. In string constructions of brane world models
 there can  be moduli whose interactions are not Planck suppressed but by a scale\footnote{ The scale $\Lambda$
 can be lower than the Planck scale for the modulus which parametrises the size of the cycle that the branes wrap. In this case $\Lambda$ is the string scale (see for e.g. \cite{other}). We note that there is a large difference between the string and Planck scale
 only if the volume of the compactification is large.} $\Lambda$. If such a modulus $(\chi)$ is the last to decay
 the CMP bound \cite{cmp,ccmp,cmmp} gets modified to 
The reheat temperature after the decay of a modulus is given by
$T_{\rm reheat} \sim  \sqrt{\Gamma M_{\rm pl}}$, where $\Gamma$ is the width of the modulus. The characteristic width of a modulus $\chi$ whose interactions in the four dimensional effective action are suppressed by a scale $\Lambda$ are given by
\bel{genw}
  \Gamma_{\Lambda} \approx { 16 \pi m^{3}_{\chi} \over \Lambda^2 }
\ee
 Combining the above with the requirement of a sufficiently high   reheat temperature for nucleosynthesis one arrives
 at a generalisation of the CMP bound \cite{cmp,ccmp,cmmp} discussed in the introduction
 \bel{newcmp}
     m_{\chi} \gtrsim   \eta^{2/3}. 30 \spa \textrm{TeV}
 \ee
 where $\eta =  {\Lambda / M_{\rm pl}} $. 

    Following the  same steps as in the earlier part of the section (while using the  lifetime of the modulus to be as given by \pref{genw}),  one can obtain the modification of our bound
 \bea
 \label{bondn}
 m_{\varphi}\gtrsim   \sqrt{16 \pi} M_{\rm pl} \eta Y^2 ~e^{-3\left(55.43  - N_k   + { 1 \over 4 } \ln ( {{ \rho_{\rm k} \over \rho_{\rm end} } })+\frac{1}{4}\ln r\right)}.
\eea
We note that both \pref{newcmp} and \pref{bondn}  scale as a positive power of $\eta$. Carrying out
a analysis as above, one  easily sees that our
bound is stronger in a large range of the phenomenologically interesting parameter space.

   Finally, we would like to  briefly discuss the cases in which the modulus primarily decays to massive particles. 
Such decay products can be super partners of standard model particles or additional Higgses.
 For models in which the primary mode of decay is  to massive particles and the lifetime scales as $m_{\varphi}^{p}$  with $p \leq -1$ our analysis will provide a lower bound on moduli masses. The bound might involve the mass of the decay products  (expression for the bound will in general be different from that given in Eq.~\eqref{bond1} and Eq.~\eqref{bond2}). In a large number of situations, the lifetime has the same form as
 \pref{width} or has the form (see for e.g \cite{darkr, darkrr, endo})
\bel{newton}
    \tilde{ \tau }\approx { 16 \pi M_{\rm pl}^2 \over m_{\varphi} \tilde{m}^2}
\ee
(i.e. $p = -1$) where $\tilde{m}$ is the mass of the decay products. Again, following 
the same steps as in the earlier part of the section one obtains a bound on the
mass of the decay products (in the case that the lifetime takes the form \pref{newton}) or a bound on the modulus
mass (in the case that the lifetime takes the same form as \pref{width}). But, a bound on the mass of the decay products translates to a bound on the mass of the modulus, as the mass of the modulus has to be heavier than the mass of the decay products. Thus, the bound  \pref{bond} applies equally well for these situations (with $p=-1$). On the other hand, for $p>-1$ our analysis will provide an upper bound for moduli masses (in terms of the mass of the decay products). This can be very interesting, although such models are not generic. We leave the detailed study of specific models for future work. In addition, thermal bath produced by the reheating may have some effects on the decay rate of the moduli \cite{Drewes:2013iaa}, thus on our bound. We will address the issue in future work.

\section{Conclusions}
\label{conclu}

We have considered cosmologies in which density perturbations are generated by the quantum fluctuations of the inflaton field at early times; the late time dynamics involves a modulus which first dominates the energy density of the universe and then decays to reheat the visible sector. In this context we have
obtained a relationship between the  mass of modulus, broad characteristics 
of post-inflationary reheating and inflationary observables (or parameters in the inflaton potential).
Together with the bound $w_{\rm re} < 1/3$ on the equation of state during reheating  (or if reheating is almost instantaneous) and 
the generic expectations on the initial displacement of the modulus at the end of inflation $( \varphi_{\rm in}
\sim M_{\rm pl})$ the relation gives a bound on the minimum mass of the modulus, see Eq. \eqref{bond}. For a large class of  inflationary models
 the bounds obtained (for values of
$n_s$ at the central value of {\small{PLANCK}}) are much stronger than the  
``cosmological moduli problem" bound. The bound is exponentially sensitive to $N_{\rm k}$ and future observations will play an important role in sharpening the bound. In fact, with precise measurements of inflationary observables it is possible that the bound becomes highly constraining.

A typical value of $N_{\rm k} \approx  50$ suggests a very high value for the modulus mass in the context of late time modulus dominated cosmology. Given a particular model of inflation, $N_{\rm k}$ is known in terms of observable parameters. Larger the value $N_{\rm k}$, more severe is the bound. 
In the case of instantaneous reheating, the bound becomes an equality and gives a prediction
for the mass of the modulus. Even
if we take $w_{\rm re} > 1/3$ (an exotic reheating phase); the value of the modulus mass obtained from our analysis (for $N_k \approx 50$) is much higher than the CMP bound
as long as the number of e-foldings during reheating are not comparable to the number of e-foldings during
modulus domination (as is expected for a light modulus). 

 The bound should  have broad implications for string/supergravity models where it is typical to have scalars interacting with Planck suppressed interactions. It can shed light on the scale of supersymmetry
breaking in the context of gravity mediated breaking; where the scale of soft masses can be tied to the moduli masses. The bound is exponentially sensitive to the number of e-foldings during inflation and hence provides a new
motivation for precision measurements of the spectral tilt.



%
\section*{{Acknowledgements}}
KD would like to thank the Harish Chandra Research Institute, the Max Planck Institute for Physics, Munich and the Abdus Salam International Centre for Theoretical Physics, Trieste for hospitality where parts of the work were done. KD is partially supported by a Ramanujan fellowship and a DST-Max Planck Society visiting fellowship. AM is partially supported by a Ramanujan fellowship.
%


\end{document}